\documentstyle[epsfig]{mn}
\begin{document}

\title
[{\it Planck Surveyor} and gravitational lenses] 
{The {\it Planck Surveyor} mission and gravitational lenses}
\author
[A. W. Blain]
{
A. W. Blain\\
Cavendish Laboratory, Madingley Road,
Cambridge, CB3  0HE.
}
\maketitle

\begin{abstract}
An extremely sensitive all-sky survey will be carried out in the 
millimetre/submillimetre waveband by the forthcoming ESA mission {\it Planck 
Surveyor}. The main scientific goal of the mission is to make very accurate 
measurements of the spatial power spectrum of primordial anisotropies in the 
cosmic microwave background radiation; however, hundreds of thousands of 
distant dusty galaxies and quasars will also be detected. These sources are 
much more likely to be gravitationally lensed by intervening galaxies as 
compared with sources discovered in surveys in other wavebands. Here the 
number of lenses expected in the survey is estimated, and techniques 
for discriminating between lensed and unlensed sources are discussed. A
practical strategy for this discrimination is presented, based on exploiting the 
remarkable sensitivity and resolving power of large ground-based
millimetre/submillimetre-wave interferometer arrays. More than a thousand 
gravitational lenses could be detected: a sample that would be an extremely 
valuable resource in observational cosmology. 
\end{abstract}  


\section{Introduction}

The properties of the population and the individual appearance of distant 
galaxies can both be modified significantly by the gravitational lensing effect of 
foreground masses. Gravitational lensing has been investigated in most detail in 
the near-infrared/optical and radio wavebands, in which faint distant galaxies 
can be observed with sub-arcsecond resolution. However, there are excellent 
prospects for extending these studies into the submillimetre waveband (Blain 
1996, 1997a,b, 1998). The fraction of 
gravitational lenses expected in a sample of galaxies selected in this waveband 
can be up to three orders of magnitude larger as compared with a sample 
selected in other wavebands.

The concept of magnification bias provides a useful way of describing the effects 
of lensing on a population of distant galaxies (Borgeest, Linde \& Refsdal 1991). If 
the surface density of galaxies with flux densities greater than $S_\nu$ per unit 
redshift is $n(S_\nu, z)$ at redshift $z$, then magnification by a factor $A$ due to 
gravitational lensing would predict a modified count, 
%\begin{equation}
$n'(S_\nu, z) = A^{-1} \, n[S_\nu A^{-1}, z].$ 
%\end{equation}
If $n \propto S_\nu^\alpha$ locally, then 
%$n' \propto S_\nu^\alpha A^{-(1+\alpha)}$ 
the magnification bias $n'/n$ is given by $A^{-(1+\alpha)}$. Hence, if 
$\alpha < -1$ then the surface density of galaxies is increased and the 
magnification bias is positive; if $\alpha > -1$ then the surface density 
of galaxies is decreased and the magnification bias is negative. 
Magnification bias was first discussed in the context of samples of bright 
quasars; however, counts of galaxies at faint flux densities in the submillimetre 
waveband are expected to be uniquely steep, with $\alpha = -3$ or less 
(Blain \& Longair 1993, 1996), and so the magnification bias for faint galaxies in 
this waveband is expected to be very significant (Blain 1996). At brighter flux 
densities the submillimetre-wave counts are expected to be less steep,
and to follow the Euclidean slope, with $\alpha = -3/2$. The magnification bias is 
expected to be largest at flux densities at which the counts begin to rise above 
the Euclidean slope. If the population of dusty star-forming galaxies evolves 
strongly, as the first detections of previously unknown galaxies in the 
submillimetre waveband appear to demonstrate (Smail, Ivison \& Blain 1997 -- 
SIB), then the largest magnification bias is predicted to occur at a flux density of 
0.1 to 1\,Jy, which is comparable 
with the sensitivity of {\it Planck Surveyor} (Bersanelli et al. 1996).

The submillimetre-wave counts of lensed galaxies are expected 
to be sensitive to changes in both world models (Blain 1998) and scenarios of 
galaxy formation and evolution. Hence, the properties of the lensed galaxies or 
quasars detected in the {\it Planck Surveyor} mission would allow the values of 
cosmological parameters to be investigated, even without detailed lens
modeling. An Einstein--de Sitter world model and Hubble's constant 
$H_0=50$\,km\,s$^{-1}$\,Mpc$^{-1}$ are assumed throughout.

\section{The properties of lenses}

The surface and flux densities of gravitational lenses in the 
millimetre/submillimetre wavebands can be predicted by combining models of 
the population of distant dusty star-forming galaxies (Blain \& Longair 1996)
with a model of the magnification distribution due to lensing as a function of 
redshift (Peacock 1982; Pei 1993). The magnification distribution can be derived 
from the mass distribution of galaxies, and takes the form $a(z) A^{-3}$ if $A$ is 
large. Estimates of $a(z)$ are presented in Fig.\,1 for both evolving and 
non-evolving models of the distribution of lensing masses. In the evolving model 
the mass distribution of the population of lensing galaxies is derived using the 
Press--Schechter formalism for structure formation by hierarchical clustering 
(Press \& Schechter 1974), in which galaxies typically become smaller and more 
numerous at larger redshifts. The probability of lensing is predicted to be smaller 
in the evolving model as compared with the non-evolving model. The form of 
$a(z)$ assumed in this letter is normalised to match the predictions of Pei (1993).

\begin{figure}
\begin{center}
\epsfig{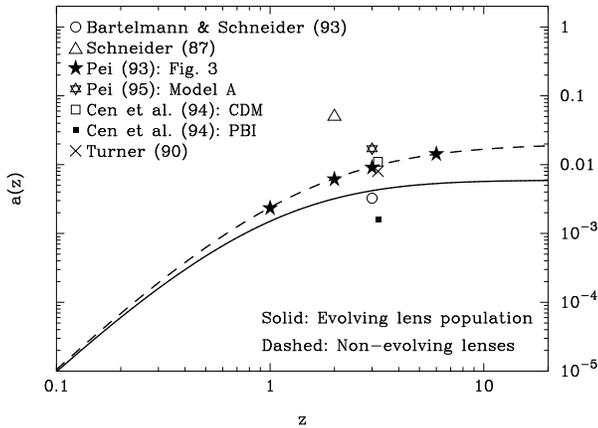}
\end{center}
\caption{The probabilities of galaxy--galaxy lensing as a function of redshift
assumed in this paper. The lensing optical depth to lensing with a
magnification greater than two out to reshift $z$ is $a(z)/8$. For 
references see Blain (1996).}
\end{figure}

Investigations of the evolution of galaxies and active galactic nuclei (AGN) in
many different wavebands (Hewett, Foltz \& Chaffee 1993; Dunlop \& Peacock 
1990; Oliver, Rowan-Robinson \& Saunders 1992; Lilly et al. 1996) indicate that 
the observed evolution of both the global star-formation rate and the luminosity 
density of AGN is consistent with pure luminosity evolution (PLE) of the form 
$(1+z)^3$ out to $z \sim 2$. The observations of SIB can be fit well by a model of 
PLE of the 60-$\mu$m luminosity function of {\it IRAS} galaxies (Saunders et al. 
1990) with the form of $(1+z)^3$out to $z=2.6$, and a constant luminosity 
evolution factor of 46.7 in the interval $2.6 \le z < 7$. The predictions of this 
model, which is consistent with other recent observations by Kawara et al. (1997) 
and Wilner \& Wright (1997) are shown in Fig.\,2, for both an evolving and a 
non-evolving distribution of lenses.

The effects of different world models are discussed by Blain (1997a, 1998) 
discusses the effects of different world models; a non-zero cosmological 
constant increases the predicted surface density of lensed galaxies as compared 
with the Einstein--de Sitter model, because both the probability of lensing at a 
particular redshift is increased and a stronger form of evolution is required in 
order to explain SIB's observations, as the smaller volume element at large 
redshifts. 

The wavelength dependence of both the lensed and unlensed counts is 
presented in Fig.\,3. The principal feature is the large increase in the ratio of the 
surface densities of lensed and unlensed galaxies at flux densities of between 0.1 
and 1\,Jy, which correspond to the onset of the steep rise in the unlensed counts 
above the Euclidean slope at these wavelengths. This increase is by almost two 
orders of magnitude as compared with the 175-$\mu$m counts, which have a form
most similar to counts expected in the optical or radio wavebands. Another 
notable feature is the predicted increase in the surface density of unlensed 0.5-Jy 
galaxies as the observing wavelength decreases. This increase is predicted to be 
larger by a factor of about 15 between wavelengths of 850 and 350\,$\mu$m. The 
counts of lensed 0.5-Jy galaxies also increase with decreasing wavelength, from 
1\,mm to about 200\,$\mu$m; however, at wavelengths shorter than about 
200\,$\mu$m this effect is reversed, and the counts are predicted to decrease as 
the observing wavelength decreases. 

\begin{figure}
\begin{center}
\epsfig{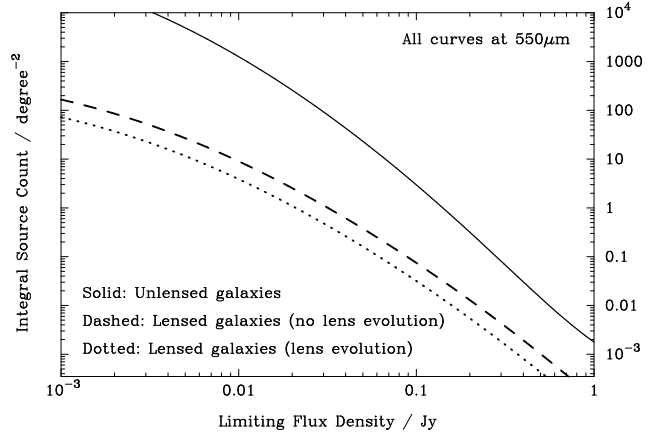}
\end{center}
\caption{The counts of lensed and unlensed field galaxies expected with and 
without evolution of the lens population.}
\end{figure} 

\begin{figure*}
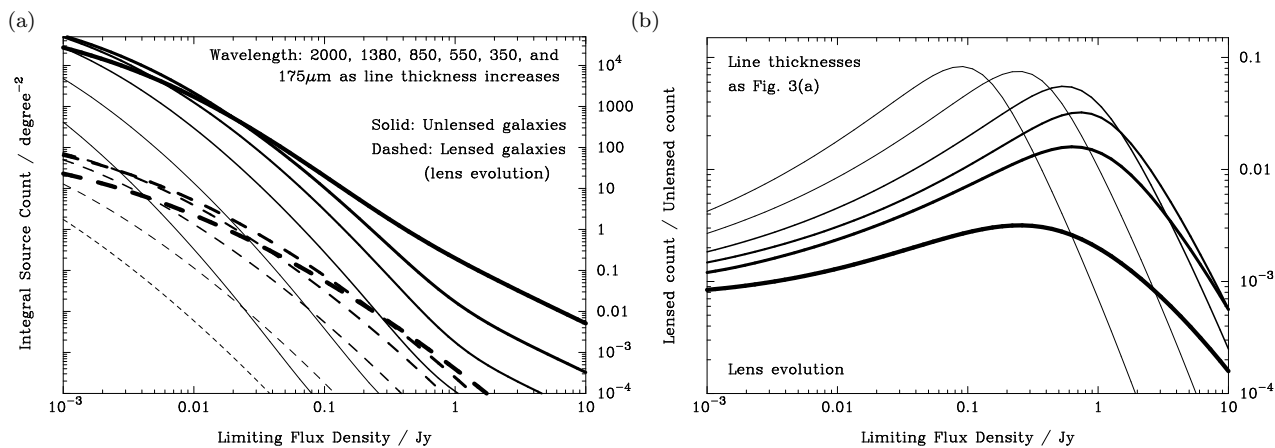

\begin{minipage}{170mm}
(a) \hskip 78mm (b)
\begin{center}
\epsfig{file=Fig3a_Planck.ps, 
width=4.95cm,bbllx=315,bblly=30,bburx=752,bbury=474}
\epsfig{file=Fig3b_Planck.ps,
width=4.95cm,bbllx=5,bblly=30,bburx=443,bbury=474}
\end{center}
\caption{The wavelength dependence of the counts of lensed and unlensed 
galaxies at the wavelengths of the {\it Planck Surveyor} HFI instrument 
passbands and at 175\,$\mu$m , assuming an evolving lens population.}
\end{minipage}
\end{figure*} 

\section{A lens survey with {\it Planck Surveyor}} 

Estimates of the numbers of galaxies that could be detected in an all-sky survey 
are listed in Table\,1, based on the sensitivities quoted by Bersanelli et al. (1996) 
and the counts discussed above. In order to produce conservative estimates, an 
evolving lens population is assumed. Note that the predicted number of lenses 
would be increased by a factor of about 5 if both a non-evolving lens model and 
a non-zero cosmological constant were assumed (Blain 1998).

\begin{table*}
\begin{minipage}{175mm}
\caption{
The sensitivities, source confusion limits and number of detections $N$ expected 
in a {\it Planck Surveyor} survey. $\sigma_{\rm cirrus}$ and $\sigma_{\rm s}$ are 
the expected Galactic cirrus confusion noise (Helou \& Beichman 1990) and 
survey sensitivity (Bersanelli et al. 1996) respectively. $\sigma_{\rm cirrus}$ is 
calculated assuming that the mean Galactic background intensity at a 
wavelength of 100\,$\mu$m is 5\,MJy\,sr$^{-1}$. $S_{\rm conf}$ is the flux density 
at which the surface density of galaxies is expected to exceed 0.03\,beam$^{-1}$; 
the expected point source confusion noise 
$\sigma_{\rm conf} \simeq S_{\rm conf} / 3$, see Blain, Ivison \& Smail (1998). An 
evolving population of lensing galaxies is assumed. In a map with Gaussian 
noise, about $10^5$, 2 and 0 pixels in the survey at 350, 550 and 850$\mu$m 
would be expected to exceed flux densities of 3, 5 and 10$\sigma$ due to 
random fluctuations.
}
{\vskip 4pt}
\begin{tabular}{ p{1.5cm}p{1.2cm}p{1.2cm}p{1.7cm}p{3.3cm}p{3.1cm}p{3.1cm} }
$\lambda$ / & $\sigma_{\rm cirrus}$ / & $\sigma_{\rm s}$ / & 
$S_{\rm conf}$ / & $N$(3$\sigma_{\rm s}$) / ($4\pi$\,sr)$^{-1}$ & 
$N$(5$\sigma_{\rm s}$) / ($4\pi$\,sr)$^{-1}$ & 
$N$(10$\sigma_{\rm s}$) / ($4\pi$\,sr)$^{-1}$\\
\end{tabular}
\begin{tabular}{
p{1.5cm}p{1.2cm}p{1.2cm}p{1.6cm}p{1.2cm}p{1.7cm}p{1.2cm}p{1.5cm}
p{1.2cm}p{1.5cm} }  
$\mu$m&  mJy & mJy & mJy & 
unlensed & lensed & unlensed & lensed & unlensed & lensed \\
\noalign{\vskip 4pt}
350  & 8.2 & 26 & 120 & 
$8.0 \times 10^5$ & $5.0\times10^3$ & $2.0\times10^5$ & $1.7\times10^3$ & 
$2.7\times10^4$ & 330 \\
550  & 7.2 & 19 & 83 &
$6.3\times10^5$ & $4.8\times10^3$ & $1.4\times10^5$ & $1.4\times10^3$ & 
$1.5\times10^4$ & 250 \\
850  & 5.5 & 16 & 40 & 
$1.4 \times 10^5$ & $1.6\times10^3$ & $2.6\times10^4$ & 420 & 
$2.2\times10^3$ & 62 \\
1380 & 3.7 & 11 & 18 & 
$9.4\times10^3$ & 190 & $1.4\times10^3$ & 43 & 96 & 5.0 \\
2000 & 2.3 & 11 & 7.3 & 
120 & 6.0 & 16 & 1.1 & 1.3 & 0.1 \\ 
\end{tabular}
\end{minipage}
\end{table*}

Between about 0.6 and 5\% of the point sources detected by {\it Planck Surveyor}
are expected to be lensed by a foreground galaxy. Despite a relatively large ratio 
of lensed to unlensed galaxies at wavelengths of 2000, 1380 and 850\,$\mu$m, the 
surface density of detectable sources is expected to be relatively small at these
wavelengths, as compared with the shorter wavelength passbands at 350 and 
550\,$\mu$m. The most useful wavelengths for detecting a sample of lenses are 
hence 350 and 550\,$\mu$m. More than a hundred lenses could be detected per 
unit solid angle in this systematic survey, and so even if about a third of these 
lenses are masked by Galactic emission at Galactic latitudes less than about 
20\,degrees, many hundreds of lenses could still be detected in the survey. This
would increase the number of known galaxy--galaxy lenses by an order of 
magnitude. The bright counts of unlensed galaxies in the 
submillimetre/far-infrared waveband would also be determined very accurately in 
such a survey, and so tight limits could be imposed to the form of evolution of 
the global star-formation rate at moderate redshifts.

\section{The discrimination of lensed and unlensed galaxies}

How efficiently could a thousand lensed sources be separated from a hundred 
thousand unlensed galaxies detected at 5$\sigma$ significance in the survey? {\it 
Planck Surveyor} will execute a multi-band survey, and so the colours of 
the detected sources could probably be used to reduce the size of the sample 
by a small factor. However, none of the sources will be resolved, and their 
positions will only be known to an accuracy of about $\pm2$\,arcmin. Other 
planned facilities -- ESA's {\it Far-Infrared and Submillimetre Space Telescope
(FIRST)} (Pilbratt 1997) and large ground-based millimetre/submillimetre-wave
interferometer arrays (MIAs; Brown 1986; Downes 1996) -- will allow this task 
to be completed in a practical amount of observing time. A strategy for following 
up the {\it Planck Surveyor} point source catalogue is outlined below. 

\subsection{Pre-selection using {\it Planck Surveyor} colours}

Colour--colour and colour--magnitude diagrams for the sources detected in a 
simulated 100-deg$^2$ sub-field of the survey are presented in Fig.\,4. The 
distribution of lensed and unlensed galaxies are clearly different in each plot: 
lensed galaxies are found at large redshifts, and if the dust temperature in 
star-forming galaxies is correlated with their luminosity, then lenses would be 
expected to have redder colours as compared with unlensed galaxies at the 
same flux densities. The field of points in the colour--colour plot (Fig.\,4a) is 
bounded at small and large redshifts by lines with a gradient of about 0.6, which 
reflects the relative wavelengths of the three relevant observing bands. If the 
emissivity of dust grains was independent of wavelength, then both lensed and 
unlensed galaxies would lie on a single line with this slope, at positions 
determined by their redshift and dust temperature. The points in Fig.\,4(a) are 
spread within a box-like region because a wavelength-dependent emissivity is 
assumed. These plots can be used to determine which sources are probably at 
low redshifts and to prioritize the sample to enhance the probability of detecting
lensed sources earlier in the follow-up work; the subsequent follow-up stages 
should proceed through the sample working from the bright red sources to
the faint blue ones. 

\begin{figure*}
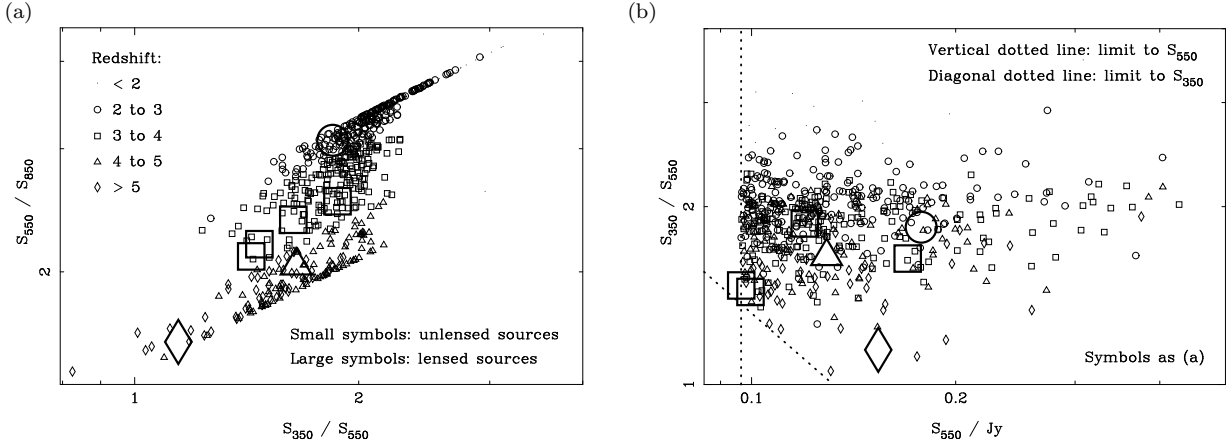

\begin{minipage}{170mm}
(a) \hskip 78mm (b)
\begin{center}
\epsfig{file=colcol_MN_rev.ps, 
width=4.95cm,bbllx=315,bblly=30,bburx=752,bbury=474}
\epsfig{file=colmag_MN_rev.ps, 
width=4.95cm,bbllx=5,bblly=30,bburx=443,bbury=474}
\end{center}
\caption{Colour--colour and colour--magnitude diagrams for both lensed and 
unlensed galaxies in a simulated 100-deg$^2$ sub-field of the {\it Planck 
Surveyor} survey. The flux densities of the plotted galaxies exceed the 5$\sigma$ 
550-$\mu$m sensitivity of the survey; in (b) they also exceed the 5$\sigma$ 
sensitivity at 350-$\mu$m. Lensed galaxies are always found at redshifts greater 
than unity, and are typically redder as compared with unlensed galaxies.}
\end{minipage}
\end{figure*}

\subsection{Determining more accurate positions} 

{\it FIRST} will be equipped with a submillimetre-wave bolometer array 
receiver with a 6-arcmin field of view and a resolution of about 20\,arcsec (Griffin 
1997). This field of view is accurately matched to the resolution of the pixels
in the {\it Planck Surveyor} map. The 350-$\mu$m flux density of candidate
lenses is expected to exceed about 0.2\,Jy, Fig.\,4(b), and so each source 
could be detected at 5$\sigma$ significance and located by {\it FIRST} in a 20-s 
integration. Assuming 200 per cent overheads, follow-up observations of $10^5$ 
lens candidates would take about 70\,days of integration, a period 
comparable with the time required for a shallow {\it FIRST} galaxy survey 
(Rowan-Robinson 1997). A ground-based instrument such as SCUBA at the 
JCMT (Cunningham et al. 1994) would require four separate pointings and a total 
integration time of about 3\,hours at 850\,$\mu$m in order to detect a candidate 
lens at 5$\sigma$ significance. Hence existing ground-based facilities are not 
sufficiently sensitive to carry out this programme. 

A large MIA is expected to detect a 50-mJy 850-$\mu$m source at 5$\sigma$ 
significance in about 28\,ms, but can only image about 0.08\,arcmin$^2$ of sky 
per pointing. Hence about 250 separate pointings, and a total integration time of 
about 7\,s would be required to image the field of each {\it Planck Surveyor} pixel 
that contains a candidate lens with such an instrument. 

Either a space-borne submillimetre-wave telescope or a ground-based 
submillimetre-wave interferometer could locate all the potential lensed 
sources in the {\it Planck Surveyor} catalogue to an accuracy better than about 
20\,arcsec in about 3\,months of observing time. Observations with {\it FIRST}
have the advantage of simultaneously determining the spectral energy 
distributions of the candidates at wavelengths shorter than 450\,$\mu$m.  
 
\subsection{Diagnosis of lenses}

Once reasonably accurate positions are known for the candidates, an MIA 
can be used to image them at a large signal to noise ratio and search for
signs of lensed arcs, rings or multiple images. The relative numbers of 
0.1-Jy lenses per unit magnification are expected to be flat out to magnifications 
of about 50 (Blain 1997a; Fig.\,4), and so most detected lens should show these 
clear signatures of strong lensing. The flux densities and redshifts of 
the candidates are expected to be similar to that of the $z=2.8$ starburst/AGN 
recently discovered by Ivison et al. (1998). Several hours of integration were
required to image and take the spectrum of this source using 4-m class
optical telescopes, and so even with 8-m telescopes available it will be quite 
impossible to follow-up the {\it Planck Surveyor} catalogue in the
optical/near-infrared waveband in a reasonable time.  

In a 1-minute 850-$\mu$m integration on a 2-km baseline, an MIA could reach a 
5$\sigma$ flux density limit of about 1\,mJy at sub-0.1-arcsecond angular 
resolution. This corresponds to the detection of a lens candidate in the {\it
Planck Surveyor} catalogue at a significance of about 250. Any multiple images, 
arcs or rings in the source will be clearly resolved using such an instrument. 
Imaging $10^5$ sources in this manner would take about 70\,days of dedicated 
observations. Of course many of the 
candidates will show no signs of multiple structure, and so could be excluded 
from the list of candidates in a shorter integration. Note that the lensing galaxy is 
expected to be optically thin to submillimetre-wave radiation and so differential 
extinction of multiple images, which affects lens surveys in the optical waveband, 
will not complicate the selection effects in a {\it Planck Surveyor} survey.
Possible other selection effects are considered in the following section. 

Ultimately the catalogue of good candidates derived from the high-resolution 
MIA images will require spectroscopic confirmation in the optical/near-infrared
in order to diagnose them as lenses unequivocally. The results of the large-area 
Sloan Digital Sky Survey (Kent 1994) will probably be of great assistance in 
identifying potential lensing galaxies. 

\subsection{Potential problems and selection effects} 

Chance superpositions of sources, multiple components of an unlensed source 
and emission from the lensing galaxy could bias the selection of candidates in 
the {\it Planck Surveyor} survey and confuse the diagnosis of lenses. 

First, the counts at flux densities comparable to the 0.1-Jy selection limit are 
expected to be about 3\,deg$^{-2}$, Fig.\,2, and so the chance alignment 
of two sources of comparable brightness within about 5\,arcsec of each other, 
which could be mistaken for different components of a multiple image, is 
expected to be of the order 0.001 per cent, equivalent to less than one source in 
the whole catalogue. About 1.5 per cent of the pixels in the {\it Planck Surveyor}
survey are expected to contain more than one detectable source; however, 
these sources will typically be well separated in follow-up observations at finer 
angular resolution.   

Secondly, although interacting galaxies show clearly separate components in 
the optical waveband, high-resolution millimetre-wave observations show that 
dust radiation typically originates from a single core of order several hundred 
parsecs in size (Solomon et al. 1997). At $z=2$ this corresponds to an angular 
scale of a few times $10^{-2}$\,arcsec, a point source for the purposes of this 
investigation. Hence, because {\it Planck Surveyor} will detect only the most 
luminous distant galaxies, it is unlikely that intrinsic multiple structure in 
distant galaxies will lead to the false diagnosis of a large number of lenses. 

Thirdly, the emission from a lensing galaxy would make the detection of a lens 
candidate more likely in the {\it Planck Surveyor} survey, and complicate its 
diagnosis. However, although in the optical waveband lensed images are
typically fainter than the lensing galaxy, the situation in the submillimetre
waveband is much more similar to that in the radio waveband, in which the 
lensed images are much brighter than the lensing galaxy. This effect was
discussed by Blain (1997b) in the context of confusion between distant lensed
submillimetre-wave galaxies in the field of a cluster and the cluster galaxies 
themselves. In Fig.\,5 the flux 
density--redshift relations expected for an $L^*$ {\it IRAS} galaxy, with a 
luminosity of about $6 \times 10^{10}$\,L$_\odot$ at the present epoch, and a 
detectable lens candidate, assumed to be at $z>0.5$ are compared. Unless the 
lensing galaxy is at $z \le 0.1$, which seems rather unlikely for the typical lens 
candidate at $z\gg 1$, the flux density from the lensed images is expected to 
dominate that of the lens, even if the lensing galaxy was a luminous {\it IRAS} 
galaxy. Most lenses would be expected to be quiescent elliptical galaxies with 
much smaller far-infrared luminosities. If a severe colour-based pre-selection 
was imposed, then perhaps contamination by the lensing galaxy could
introduce a small bias into the 
sample; however, we have shown that the follow-up of the {\it Planck Surveyor} 
survey is still practical even without any pre-selection.
%Very deep follow-up observations of the best lensing candidates, with 
%confusion-limited hour-long integrations using an MIA (Blain et al.\ 1998) 
%would reach about an order of magnitude deeper than the limit shown in 
%Fig.\,5, and so a dusty lensing galaxy with a star-formation rate in excess of 
%about 1\,M$_\odot$\,yr$^{-1}$ should be detectable, allowing purely 
%submillimetre-wave confirmation of the lensing properties. 

\begin{figure}
\begin{center}
\epsfig{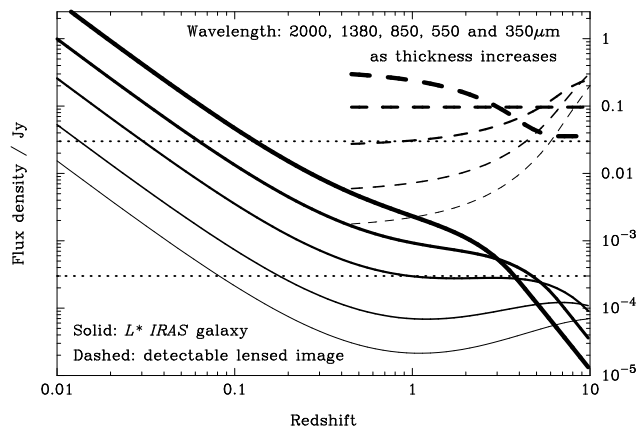}
\end{center}
\caption{The flux density--redshift relations expected for an $L^*$ {\it IRAS}
galaxy and a lensed object detected at 5$\sigma$ at 550\,$\mu$m in a {\it Planck 
Surveyor} survey. A low dust temperature of 30\,K is assumed to boost the 
low-redshift flux densities. The upper and lower dotted lines correspond to the 
approximate 1$\sigma$ sensitivities of the {\it Planck Surveyor} survey, Table\,1, 
and a 1-minute integration with a large MIA respectively.}
\end{figure} 

\section{The optimal strategy} 

It will be practical to compile a sample of about a thousand lenses, starting from 
the $10^5$ 5$\sigma$ detections expected in the {\it Planck Surveyor} point 
source catalogue, by conducting a concerted programme of observations using 
{\it FIRST}, large MIAs and conventional optical telescopes lasting several 
months. The most promising candidates can be pre-selected for follow-up 
observations based on their submillimetre-wave colours in the catalogue. 
{\it FIRST} or an MIA can then be used to determine the positions of the 
candidates with sufficient accuracy to permit a programme of high-resolution 
follow-up imaging observations using an MIA. Any lensed arcs, rings and 
multiple images in the candidate sources will be identified at this stage. Finally, 
the remaining sample of about a thousand excellent candidates would be 
subjected to spectroscopic confirmation or refutation as lenses using 8-m class 
telescopes in the optical/near-infrared wavebands. 

\section{Conclusions}

The {\it Planck Surveyor} mission has the potential to detect a very large sample
of gravitationally lensed galaxies and quasars. The exact size of the sample
will depend on the properties of distant dusty galaxies and quasars, the 
cosmological parameters, the observing wavelength and the sensitivity of the 
survey, but about a thousand lenses and about a hundred thousand unlensed 
galaxies are expected to be detected a 5$\sigma$ significance at wavelengths of 
350 and 550\,$\mu$m. A practical strategy for submillimetre-based follow-up
observations has been outlined using {\it FIRST} and a large ground-based 
millimetre/submillimetre-wave interferometer array. The scientific rewards from 
the compilation of such a large unbiased sample of lenses would be very 
considerable, and it is difficult to see how such a sample could be compiled in 
any other waveband. 

\section*{Acknowledgements}

I would like to thank an anonymous referee for his/her very helpful comments
that clarified Sections\,3 and 4 significantly. This work has benefited
greatly from observations made using the James Clerk Maxwell Telescope in 
collaboration with Ian Smail, Rob Ivison and Jean-Paul Kneib.

\end{document}